# Senescent Cells in Growing Tumors: Population Dynamics and Cancer Stem Cells

Caterina A. M. La Porta[1], Stefano Zapperi[2,3]*, James P. Sethna[4]

1 Department of Biomolecular Science and Biotechnology, University of Milano, Milano, Italy, 2 CNR-IENI, Milano, Italy, 3 ISI Foundation, Torino, Italy, 4 LASSP, Department of Physics, Cornell University, Ithaca, New York, United States of America

## Abstract

Tumors are defined by their intense proliferation, but sometimes cancer cells turn senescent and stop replicating. In the stochastic cancer model in which all cells are tumorigenic, senescence is seen as the result of random mutations, suggesting that it could represent a barrier to tumor growth. In the hierarchical cancer model a subset of the cells, the cancer stem cells, divide indefinitely while other cells eventually turn senescent. Here we formulate cancer growth in mathematical terms and obtain predictions for the evolution of senescence. We perform experiments in human melanoma cells which are compatible with the hierarchical model and show that senescence is a reversible process controlled by survivin. We conclude that enhancing senescence is unlikely to provide a useful therapeutic strategy to fight cancer, unless the cancer stem cells are specifically targeted.

Funding: JPS acknowledges NCI-U54CA143876 for support. CAMLP is supported by PRIN 2008BP25KN004. The funders had no role in study design, data collection and analysis, decision to publish, or preparation of the manuscript.

Competing Interests: The authors have declared that no competing interests exist.

* E-mail: stefano.zapperi@cnr.it

## Introduction

Cancer cells are characterized by their persistent proliferation, but just as for normal cells [1] tumor cells can go senescent, halting their growth [2,3]. The molecular basis for the induction of senescence appears to be a combination of several mechanisms such as telomerase shortening, DNA-damage and oxidative stress [3]. It has been suggested that senescence should be present only in pre-neoplastic cells [3] but there is evidence that senescence markers increase during tumor progression [4]. This is puzzling since it is usually assumed that tumors can only grow if senescence is avoided. Recent experiments are challenging this conventional view of cancer, showing that only a small fraction of cancer cells, the cancer stem cells (CSCs) [5], actively drive tumor growth. The implications of this finding for tumor cell senescence have hitherto not been explored.

According to the CSC hypothesis, tumors behave in analogy with normal tissues, whose growth is controlled by a small population of slowly replicating stem cells with the dual capacity of either self-renewal or differentiation into the more mature cells required by the tissue. The crucial difference between tissue stem cells and CSCs lies in their proliferation properties. Stem cells in tissues tend to keep their number constant either deterministically by enforcing asymmetric division, or stochastically by balancing proliferation and depletion probabilities [6,7]. This restriction does not hold for CSCs. The role of CSCs has important implications for therapeutic approaches. According to the conventional view, the success of a treatment is measured by the number of tumor cells killed; in contrast, according to the CSC hypothesis, only the CSC subpopulation matters in the end for complete eradication. First evidences for the existence of CSCs came from hematological tumors [5] and later from solid tumors such as breast cancer [8] and melanoma [9–16]. The presence of CSCs in melanoma is currently debated. It was argued that to obtain reliable estimates of the number of tumor initiating cells one should use highly immunocompromised mice for tumor xenografts [17]. Different groups, however, reported conflicting results with slight differences in assay conditions and mouse models [14,17]. Furthermore, several putative CSC markers appears to be reversibly expressed in melanoma [18] and in breast cancer [19].

In this paper, we analyze cell senescence during the growth of melanoma cells both in vitro and in tumor xenografts. We find that the fraction of senescent cells increases considerably after a few months of cultivation, slowing down the growth of the cell population. This process is, however, only transient: after some time senescence almost disappears and growth resumes at the initial rate. We also show that senescence is a reversible process controlled by survivin: by overexpressing survivin in senescent cells, we are able to decrease senescent markers and increase cell proliferation. These results can be interpeted in terms of the hierarchical cancer model where only CSCs replicate indefinitely and senescence reflects the loss of proliferative capacity of other cancer cells. To prove this point we sort cancer cells according to a putative CSC marker, ABCG2, and show that senescence is more prevalent for negative cells, suggesting that CSC are able to rejuvenate an otherwise senescent cell population.

To understand quantitatively the experimental results, we propose a mathematical model for cancer growth that is compatible with the existence of CSCs in melanoma, provides an unambiguous interpretation of experimental data, and explains quantitatively the occurrence of senescence in tumors.

**Author Summary**

It is commonly believed that cell senescence – the loss of replicative capacity of cells – acts as a barrier for tumor growth. Here we follow the evolution of senescence markers in melanoma cells and find that while most cancer cells eventually turn senescent, this is at root irrelevant for the long-term growth rate of a tumor. To demonstrate this, we construct a mathematical population dynamics model incorporating cancer stem cells which is able to reproduce quantitatively the experimental data. Our results support the existence of cancer stem cells in melanoma and explain why it is difficult to fight cancer by inducing senescence in cancer cells. Only a fraction of the cells are susceptible to senescence, but those cells are irrelevant for tumor growth. A successful therapeutic strategy should instead target cancer stem cells, which are, however, likely to be strongly resistant to drug induced senescence.

While in the stochastic cancer model, senescence is due to spontaneous mutations of tumor cells, in the hierarchical model only CSCs replicate indefinitely and senescence reflects the loss of proliferative capacity of other CCs. These observations can be recast in a mathematical framework using the theory of branching processes, a generic model for a growing population [20,21]. Branching processes were first proposed at the end of the XIX century and found wide application in physics and biology with examples ranging from nuclear reactions [20] to evolution theory [21], tissue growth [6,22–24] and cancer progression [25,26]. The main limitation of branching processes is that they do not account for interactions between cells which could be relevant for the growth of a tumor, especially in vivo. Despite this shortcoming, our model allows for a good description of the experiments, providing an indirect confirmation of the CSC hypotesis for melanoma with important implications for therapeutic strategies based on the induction of senescence in cancer.

## Results

### Hierarchical model for cancer growth

Our theory has biologically motivated ingredients characterizing the probabilities for cells to duplicate, become senescent or die: According to the CSC hypothesis, cells are organized hierarchically, with CSCs at the top of the structure. CSCs can divide symmetrically with rate $R_d$ giving rise to two new CSCs with probability $p_2$, two CCs with probability $p_0$, or asymmetrically with probability $p_1 = 1 - p_0 - p_2$ giving rise to a CSC and a CC. While CSCs can duplicate for an indefinite amount of time, CCs become senescent after a finite number of generations $M$ and then eventually die rate $q$ (Fig. 1). The model kinetics depends on the combination $\epsilon = p_2 - p_0$, the average relative increase in the number of CSCs after one duplication, rather than on $p_0$, $p_1$ and $p_2$ separately. In normal tissues, the stem cell population should remain constant which implies that $\epsilon = 0$ [6,27], while in tumors we expect $\epsilon > 0$.

The kinetics of the cell populations for this model can be solved exactly. To this end, we first derive recursion relations linking the average cell populations at each generation. Denoting by $S^N$ the average number of CSCs after $N$ generations, by $C_k^N$ the average number of CCs, where $k = 1, ..., M$ indicates the "age" of the CCs (i.e. the number of generations separating it from the CSC from which it originated), and by $D^N$ the average number of senescent cells, we obtain

$$
\begin{aligned}
S^N &= (1+\epsilon)S^{N-1} \\
C_1^N &= (1-\epsilon)S^{N-1} \\
&... \ ... \ ... \\
C_k^N &= 2C_{k-1}^{N-1} \\
D^N &= (1-q)D^{N-1} + 2C_M^{N-1}.
\end{aligned}
\tag{1}
$$

These relations are derived considering that CSCs can only originate from other CSCs either by a symmetric division – two CSCs are generated with probability $p_2$ – or by an asymmetric one, in which case a single CSC is generated with probability $1 - p_2 - p_0$. Hence each CSC generates an average of $2p_2 + 1 - p_2 - p_0 = 1 + \epsilon$ new CSCs. CSCs also generate CCs ($k=1$) by asymmetric CSC divisions, with probability $1 - p_2 - p_0$ or by symmetric CSC division with probability $p_0$, yielding an average of $1 - p_2 - p_0 + 2p_0 = 1 - \epsilon$ CCs. Two CCs are generated by duplication of other normal cells with unit probability ($k = 2...M$). Senescent cells accumulate at each generation when normal cells (with $k = M$) lose the ability to duplicate and die with probability $q$. Eqs. 2 can be solved explicitly. We consider first an initial condition with CSC only (i.e. $C_k^0 = D^0 = 0$) and obtain

$$
S^N = (1+\epsilon)^N S^0
$$

$$
C_k^N = \begin{cases} S^N \left(\dfrac{2}{1+\epsilon}\right)^{k-1} \dfrac{1-\epsilon}{1+\epsilon} & \text{for } N > k \\ 0 & \text{for } N \le k \end{cases}
\tag{2}
$$

$$
D^N = \begin{cases} S^N \dfrac{(1-\epsilon)}{\epsilon + q} \left(\dfrac{2}{1+\epsilon}\right)^M \left(1 - \left(\dfrac{1-q}{1+\epsilon}\right)^{N-M}\right) & \text{for } N > M \\ 0 & \text{for } N \le M \end{cases}
$$

Eqs. 2 implies that the average CSC population $S^N$ grows exponentially fast, and that after a large number of generations (i.e. for $N \gg M$), all cells populations, and hence the size of the tumor, are proportional to the CSC population. This result confirms that the population of CSCs is the driving force behind tumor growth. Note that the population of senescent cells is also driven by the growth of CSCs and therefore cancer growth and senescence are inextricably linked.

To interpret experimental data, it is important to consider the case in which the initial condition is composed of a mixed population of CSCs and CCs. The solution in this case can be obtained as a linear superposition of the solution of Eqs. 2 and the solution with initial conditions $S^0 = 0$, $C_k^0 > 0$ and $D^0 > 0$. For simplicity, we consider the case of a uniform distribution for the ages of CSCs: $C_k^0 = D^0 = c$, and obtain

$$
\hat{S}^N = 0
$$

$$
\hat{C}_k^N = \begin{cases} 2^k c & \text{for } N < k \\ 0 & \text{for } N \ge k \end{cases}
\tag{3}
$$

$$
\hat{D}^N = \begin{cases} c(1-q)^N + c\dfrac{2}{1+q}(2^N - (1-q)^N) & \text{for } N \le M \\ c(1-q)^N\left(1 + \dfrac{2}{1+q}\left(\left(\dfrac{2}{1-q}\right)^M - 1\right)\right) & \text{for } N > M \end{cases}
$$

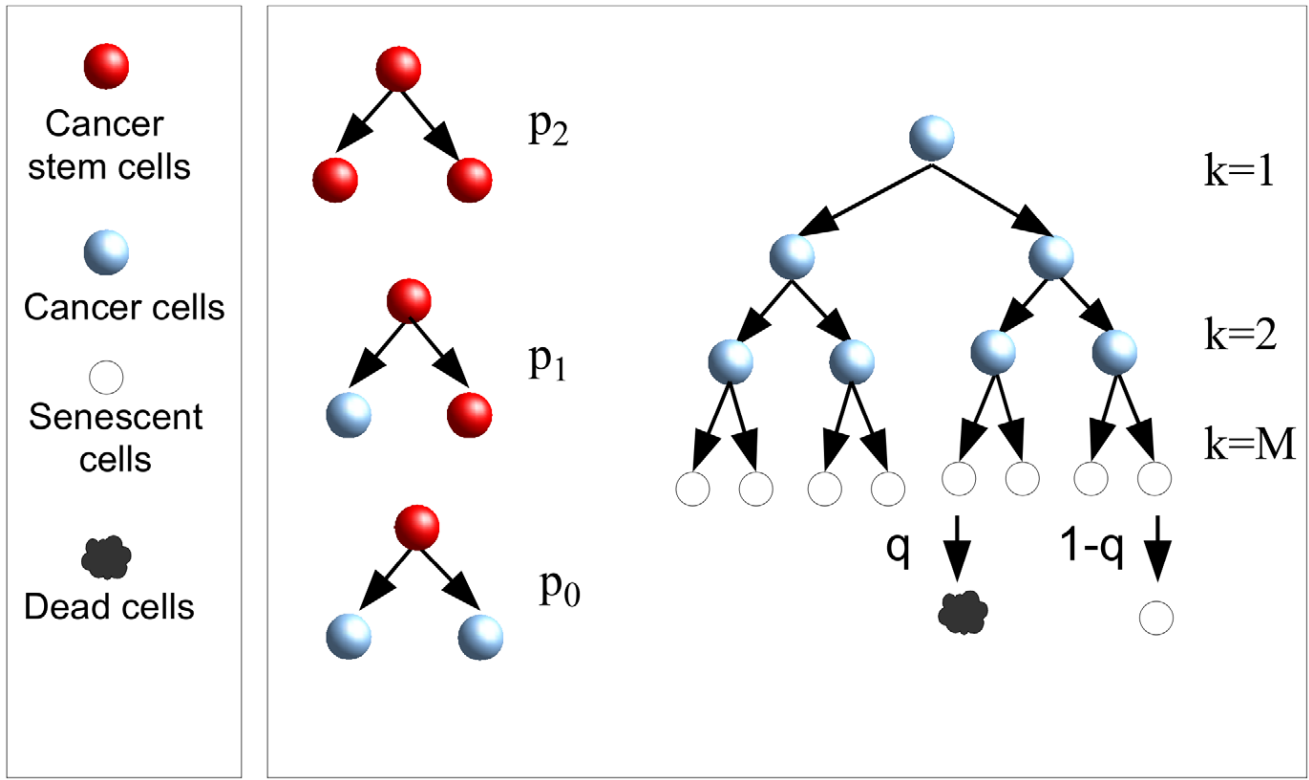

**Figure 1. Branching processes for cancer growth.** At each generation, CSCs can divide symmetrically, giving rise to two CSCs with probability $p_2$ or to two CCs with probability $p_0$, or asymmetrically with probability $p_1 = 1 - p_2 - p_0$ giving rise to a CSC and a CC. Cancer cells can only divide a finite number of times $M$ (in the figure $M = 3$), after that they become senescent. Senescent cells die with probability $q$ at each generation.
doi:10.1371/journal.pcbi.1002316.g001

The complete solution for the total number of cells, corresponding to an initial condition $S^0 > 0$ and $C_k^0 = D^0 = c$, is given by

$$n_{tot}^N = S^N + \tilde{S}^N + \sum_{k=1}^{M} (C_k^N + \hat{C}_k^n) + D^N + \hat{D}^N \qquad (4)$$

and is plotted in Fig. 2A, while in Fig. 2B we report the fraction of senescent cells.

Eqs.2 represent only a transient contribution to the cell population and does not influence the asymptotic fractions (i.e. for $N \gg M$) of senescent cells $f_{SC}^\infty$ and CSCs $f_{CSC}^\infty$ which are readily obtained dividing the results in Eqs. 2 by the total number of cells at each generation:

$$f_{SC}^\infty = \frac{1-\epsilon}{1+q} \qquad (5)$$

$$f_{CSC}^\infty = \frac{q+\epsilon}{1+q} \left( \frac{1+\epsilon}{2} \right)^M. \qquad (6)$$

The asymptotic solutions are plotted in Fig. S1 as a function of $\epsilon$ for different values of $q$ and $M$.

It is instructive to use the solution to compare how cells become senescent in normal tissues and in cancers. In the long time limit, the asymptotic fraction of senescent cells is equal to $f_S^\infty = (1-\epsilon)/(1+q)$ and is independent of the number of duplications $M$ needed to induce senescence (see Fig. S1). For normal stem cells, $\epsilon = 0$ and the percentage of senescent cells is expected to be large, reaching 100% in the limit of $q = 0$, when senescent cells never die. For cancer cells, $\epsilon > 0$ and the fraction of senescent cells is smaller but still non-vanishing. Similarly, the CSC fraction is given by $f_{CSC}^\infty = \frac{q+\epsilon}{1+q} \left( \frac{1+\epsilon}{2} \right)^M$ which is smallest for normal stem cells and increases as a function of $\epsilon$ in tumors (see Fig. S1). Aggressive tumors such as melanoma should be characterized by high values of $\epsilon$ and therefore by relatively high values of the CSC fraction. Beside asymptotic fractions, the model allows one also to characterize the evolution of the populations, i.e. the time-dependent evolution of the senescent cell fractions (see Fig. 2). The results show that the level of senescence in tumors strongly depends on the fraction of CSCs and on time.

## Growth of mesenchymal stem cells

To quantitatively confirm that the model is able to predict the growth properties of stem cells, we first compare its results with experimental data for normal stem cells, and then turn to cancers. It has been shown that long-term in vitro culture of mesenchymal stem cells (MSC) induces replicative senescence with important implication for the therapeutic application of MSC preparations [28,29]. In particular, the authors show a rapidly increasing fraction of cells that are positive to $\beta$-gal, a senescence marker; after roughly 50 days, the fraction of senescent cells reaches 80% of the population. By looking at the growth curves for cells extracted from donors of different ages, the authors could not establish a correlation between senescence and aging [29].

To validate our model, we have used it to fit the growth curves reported in Ref. [28]. In order to compare the model with the experiments, we should first express the evolution equations as a function of time rather than generation number. This is easily done introducing the rate of cell division per day $R_d$ and replacing in Eqs. 2–3 the generation index $N$ by $R_d t$, where $t$ is time expressed in days. In this way, Eq. 4 can directly be compared with the growth curves reported in Refs. [28,29]. In particular, we use Eq. 4 for $\epsilon = 0$, $q = 0$ and $c = 0$. Under this conditions the expression reduces to a much simpler form

$$n_{tot}(t) = \begin{cases} S^0 2^{R_d t} & \text{for } t \le M/R_d \\ S^0 2^M (1 + R_d t - M) & \text{for } t > M/R_d \end{cases} \qquad (7)$$

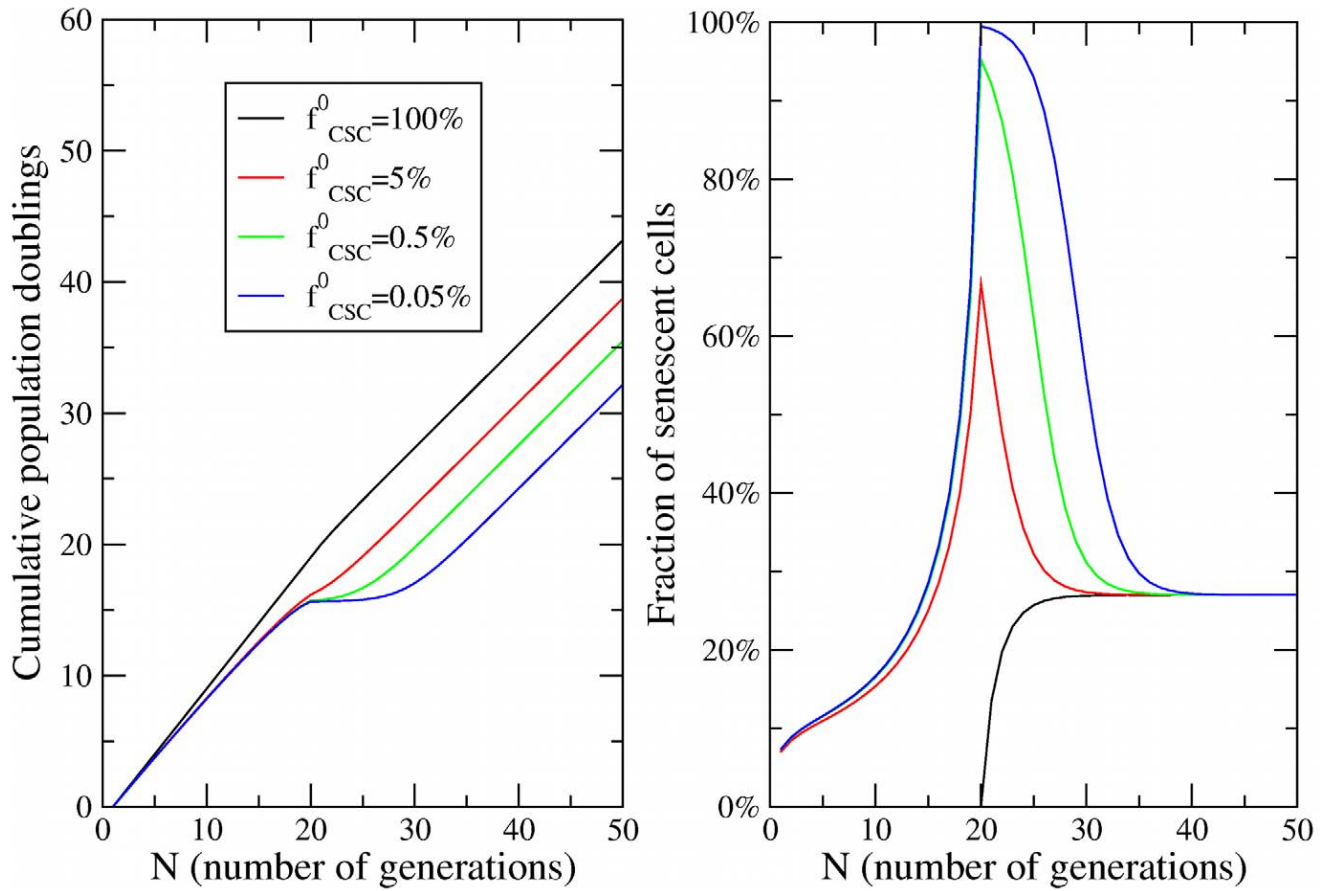

**Figure 2. Dynamics of cell fractions.** (a) The growth of the cell population obtained in the model for $\epsilon = 0.7$, $q = 0.1$, $M = 20$ and different values of the initial fraction of CSCs $f_{CSC}^0$ as a function of the number of generations $N$. (b) The evolution of the fraction of senescent cells corresponding to the same parameters.
doi:10.1371/journal.pcbi.1002316.g002

We then compute the CPD based on Eq. 14 and perform a two parameters fit in terms of $R_d$ and $M$ (see Figs. 3 and S2).

We have also analyzed similar data reported in Ref. [29] for the growth of MSC isolated from human bone marrow for the iliac crest (BM) and from the femoral head (HIP) (see Fig. 2). Although significant fluctuations between donors are revealed, the results show in general a decrease of the parameter $M$ with the age of the donor (see Fig. 4). Thus, older people have fewer generations of reproduction before senescence. The plot also shows for donors of similar age that $M$ is smaller for BM-MSC than for HIP-MSC, suggesting that the former are more prone to become senescent. On the other hand, the cell division rate fluctuates around the value of $R_d = 0.5$ divisions/day, as reported in Ref. [27], without any apparent correlation with age.

## Cancer growth and senescence

In recent publications, one of us identified aggressive subpopulations expressing two stem cell markers, ABCG2 [9] and CXCR6 [16], in human melanoma cell lines, making this tumor a suitable candidate to test our theory. Here, we analyze the level of expression of the senescence marker $\beta-$gal in three human melanoma cell lines, WM115, IGR39 and IGR37 [9,16]. IGR39 and IGR37 cells have been sorted into two subpopulations according to the expression of the ABCG2 marker [16]. In Fig. 5A and Table S1, we report the level of the $\beta-$gal senescent marker as a function of time for ABCG2 positive (ABCG2+) and negative cells (ABCG2−). In both cases, we observe an increase in the percentage of $\beta-$gal positive cells after 80–90 days of growth followed by a decrease to the initial level after 120 days (see Table S1). While the data follow a similar trend for the two subpopulations, the peak for IGR39 ABCG2− cells is much higher, reaching 90% of the cells. In order to confirm that the difference in the expression of the senescence marker is related to a different proliferative behavior for the two subpopulations, in Fig. 5B we report the long-term growth curves corresponding to IGR39. The results show that ABCG2+ cells grow substantially more than ABCG2− cells (i.e. 23 times more). Furthermore, the peak percentage of $\beta-\mathrm{gal}$ coincides in time with a marked slow down in the growth (Fig. 5). When the level of $\beta-\mathrm{gal}$ expression falls back to the initial value, growth is also restored (Fig. 5B).

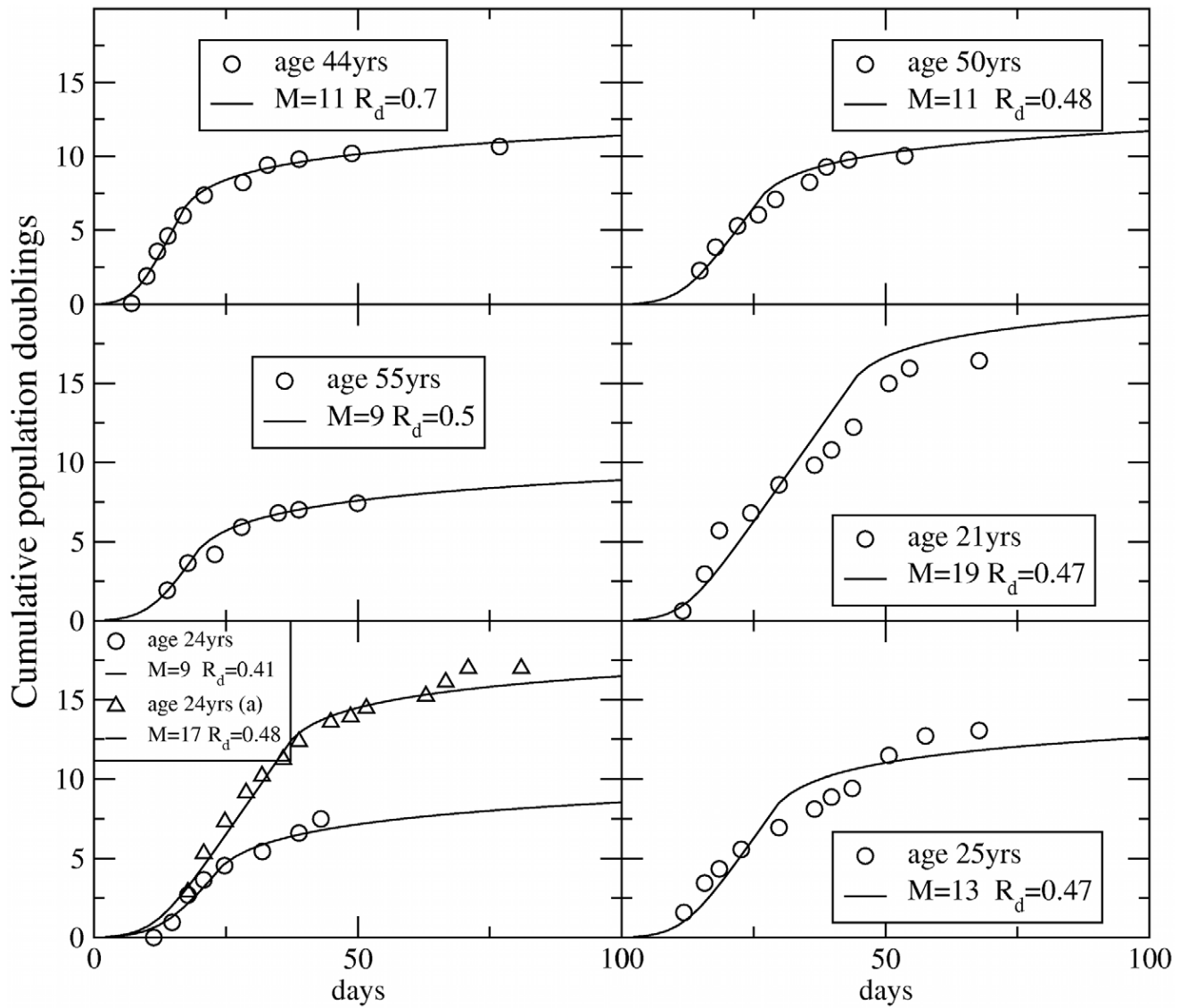

**Figure 3. Growth curves of mesenchymal stem cells (BM).** The growth of populations of MSC isolated from the bone marrow from the iliac crest in terms of cumulative population doublings are fitted by the model. Experimental data are obtained from Ref. [28]. The best fit is obtained varying $M$ and $R_d$ in Eq.7.
doi:10.1371/journal.pcbi.1002316.g003

These data can be described by our theoretical model, if we assume that ABCG2+ and ABCG2− data are only distinguished by a different initial fractions of CSCs $f_{CSC}^0$. Fitting simultaneously the growth curves and the $\beta$ gal concentrations with the model (see Fig. 5 and Materials and Methods for details), we obtain $f_{CSC}^0 = 0.63 \pm 0.05\%$ for ABCG2− and $f_{CSC}^0 = 16 \pm 1\%$ for ABCG2+ cells. Hence, our model suggests that expression of ABCG2 is strongly correlated with CSCs (almost a factor of thirty), but only a relatively small fraction of ABCG2 expressing cells are CSCs. Notice that the final CSC fraction is different from the initial one: using the fit parameters we estimate $f_{CSC}^\infty = 0.2\%$. As shown in Fig. 5A the fraction of senescent cells predicted by the model is in good agreement with the experimental data, predicting a large senescence peak for ABCG2− cells. Notice that we could easily find parameters producing a clear peak in the $\beta-\mathrm{gal}$ concentration also for ABCG2+ cells. Our multicurve fit, however, is dominated by the (higher precision) growth curves and the resulting best fit parameters yield no peak in the $\beta-\mathrm{gal}$ plot. For ABCG2− cells, the peak in the theory curves arises after $M$ generations. The experimental peak in senescence and the associated dip in doubling rate after $\sim 38$ generations directly implies a long-term memory of the initial preparation, embodied by our senescence generation $M$, a parameter that we expect may be smaller for other growth conditions and tumor types.

## Senescence in the stochastic model of cancer growth

The peak in cellular senescence is quantitatively described by the hierarchical model and it would be hard to reconcile this result with the stochastic cancer model, where cancer cells are not organized hierarchically. To illustrate this point, we reformulate the stochastic model in mathematical terms assuming that senescence can only occur as a result of random mutation. In this model, we distinguish between two cell populations, cancer cells (CC) and senescent cells. CC duplicate with probability $1-p-q$, die with probability $q$ and become senescent with probability $p$, while senescent cells do not duplicate but can die with probability $q$. The average numbers of CC $C^N$ and senescent cells $D^N$ after $N$ generations follow recursion relations

$$C^N = 2(1-p-q)C^{N-1} \tag{8}$$

$$D^N = (1-q)D^{N-1} + pC^{N-1}. \tag{9}$$

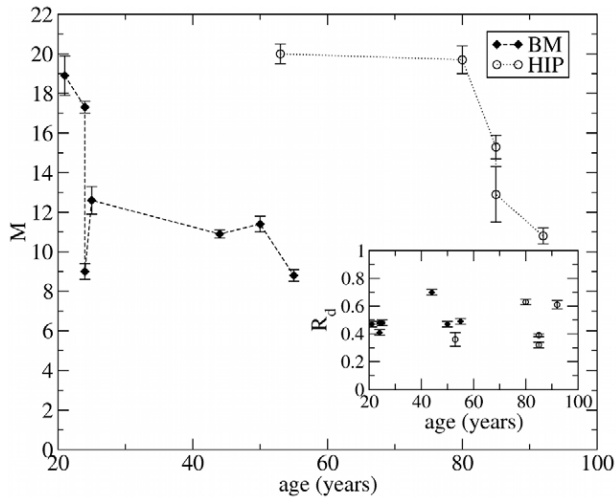

**Figure 4. Effect of the donor age on cell senescence.** The value of the parameter $M$ obtained in Fig. 3 as a function of the age of the donor for MSC isolated from bone marrow from iliac crest (BM) and femoral head (HIP). The decrease of $M$ with age indicates that for older donors cell senescence occurs more rapidly. HIP-MSC show a larger value of $M$ than MSC-BM. We report our estimates for the parameter $R_d$ in the inset, showing values compatible with $R_d = 0.5$ reported in Ref. (26) and no significant age dependence.
doi:10.1371/journal.pcbi.1002316.g004

We can solve the recursion relations explicitly obtaining:

$$C^N = [2(1-p-q)]^N C^0 \tag{10}$$

$$D^N = \left(\frac{p(1-q)}{1-2p-q}\right)([2(1-p-q)]^N - [1-q]^N)C^0, \tag{11}$$

where we have assumed for simplicity that $D^0 = 0$. The solution implies that the number of CC grows as long as $2(1-p-q) > 1$ and shrinks otherwise, yielding a condition for tumor progression. Dividing $C^N$ and $D^N$ by the total population size $C^N + D^N$, we obtain the relative fractions $f_{CC}$ and $f_{SC}$ of CC and senescent cells, respectively. In the asymptotic limit (i.e. large $N$), we obtain

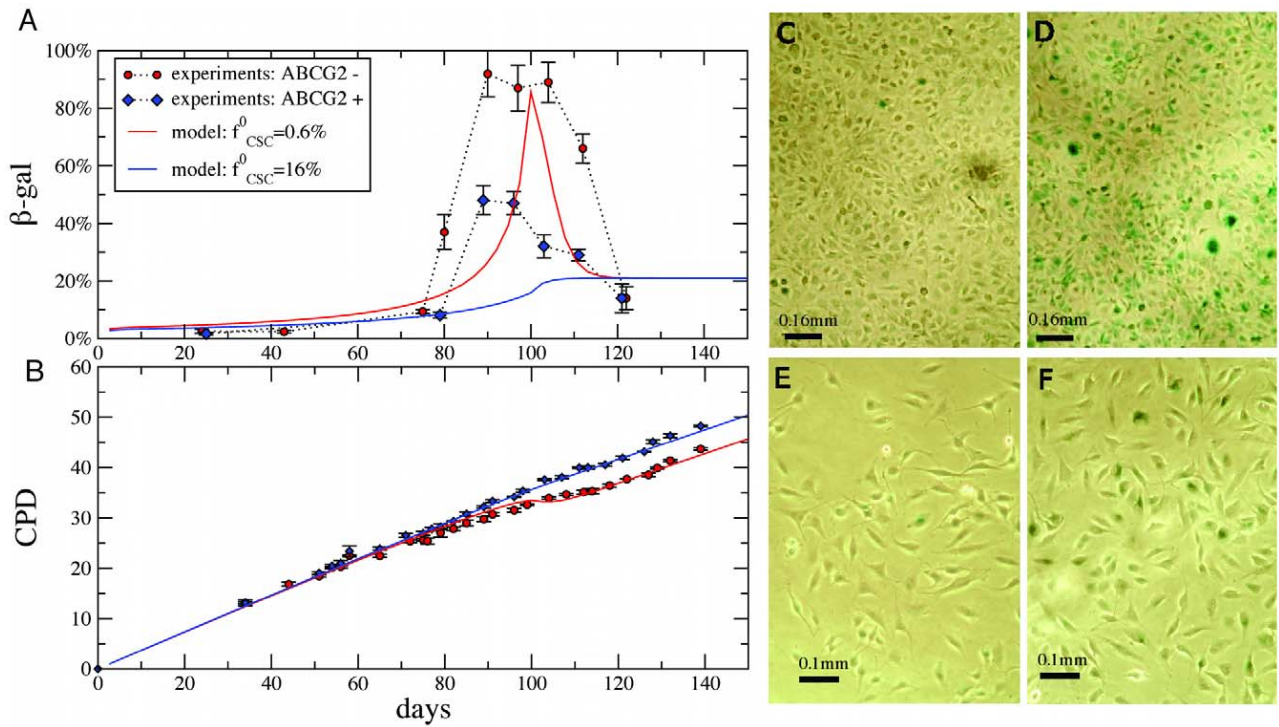

**Figure 5. Growth and senescence of IGR39 ABCG2+ and ABCG2− human melanoma cells.** Panel A shows the fraction of $\beta$-gal positive cells for ABCG2+ and ABCG2− cells. In both cases, senescence increases rapidly after 90 days; the fraction reaches higher values for ABCG2− cells. After 100 days, however, the fraction drops down to the asymptotic values. Panel B shows the corresponding long-term growth curves, compared with predictions of the model, using the best fit parameters, used also for solid lines in Panel A. Panels C–F display $\beta$-gal staining for ABCG2− (C–D) and ABCG2+ cells (E–F) at different stages: at the beginning (C and E) and at the time of the fraction peak (D and F). The fractions of $\beta$-gal positive cells (shown in green) is largest in panel D.
doi:10.1371/journal.pcbi.1002316.g005

steady-state solutions

$$f_{CC} = \begin{cases} \dfrac{1-2p-q}{1-p-q-pq} & \text{for } 1-2p-q>0 \\ 0 & \text{for } 1-2p-q<0 \end{cases} \qquad (12)$$

$$f_{SC} = \begin{cases} \dfrac{p(1-q)}{1-p-q-pq} & \text{for } 1-2p-q>0 \\ 1 & \text{for } 1-2p<0 \text{ and } q=0 \\ 0 & \text{for } 1-2p-q<0 \text{ and } q>0 \end{cases} \qquad (13)$$

The approach to the steady-state is reported in Fig. 6A and the steady-state solutions are plotted in Fig. 6B in the case $q=0$. These results show that the fraction of senescent cells should evolve rapidly towards a steady state in way that is incompatible with the experiments, providing additional evidence supporting the existence of CSCs in melanoma. The present stochastic model is clearly oversimplified and one could think of more elaborate models involving heterogeneous cell populations without a hierarchy. We do not see, however, how we can explain a peak in senescence after 90 days without assuming the presence of at least two distinct subpopulations one of which undergoes senescence after this long interval.

### Senescence in tumor xenografts

In Fig. 7, we report the evolution of the fraction of senescent cells in human melanoma WM115 cells according to the $\beta-\text{gal}$ marker. Fig. 7 shows an increase in the fraction of cells expressing this marker, reaching more than 40% after 80 days of cultivation.

We find a similar level of $\beta-\text{gal}$ also in tumor xenografts after 60 days (Fig. 7D). This confirms that our finding of senescent melanoma cells is not restricted to growth in vitro.

### Hypoxia does not affect growth of melanoma cells

In order to confirm that the observed cell senescence in xenografts (Fig. 7) is not due to hypoxia, we grow WM115 cells under normal or hypoxic conditions (see Materials and methods). Only a slight inhibition on cell growth was observed 48 hrs after incubation under hypoxic condition. This behavior is explained by the induction of the hypoxia inducible factor $-1\alpha$ (HIF$-1\alpha$) after 18 hours of hypoxic conditions (see Fig. 3). HIF$-1\alpha$ is a pleiotropic transcription factor typically activated in response to low oxygen tension as well as other stress factors in normoxic conditions. Upon activation HIF$-1\alpha$ mediates the transcriptional activation of target genes involved in a variety of processes comprising stress adaptation, metabolism, growth and invasion, but also apoptotic cell death. While we cannot exclude that some form of environmental stress factor affect our results, we conclude that hypoxia is not a likely cause for the senescence we observe in xenografts; our in vitro cells do not experience hypoxia.

### Survivin reverses senescence

Since it has been recently shown that survivin allows cells to escape senescence [30], we transfect IGR37 ABCG2− cells at the peak of $\beta-\text{gal}$ (corresponding to the 98th day for IGR37 cells, see Table S1) with a survivin-GFP or GFP alone [31]. We then quantify the resulting senescence with $\beta-\text{gal}$ activity and visualize cell proliferation with crystal violet assays (see supplement for more details). In Fig. 8A, we show that the overexpression of survivin induces a dramatic decrease of $\beta-\text{gal}$ activity. Importantly, under the same conditions survivin enhances cell proliferation as shown by the colony size distributions (Fig. 8B), the number of colonies and the fraction of area covered by colonies (see Table S2). The effect of survivin is absent for unsorted IGR37 cells when the level of $\beta-$gal is low (see Fig. 8B). Moreover, when the percentage of $\beta-$gal positive cells is below the peak the effect is very small (see Fig. 5). These data provide a further confirmation of the presence of senescent cells in growing tumors.

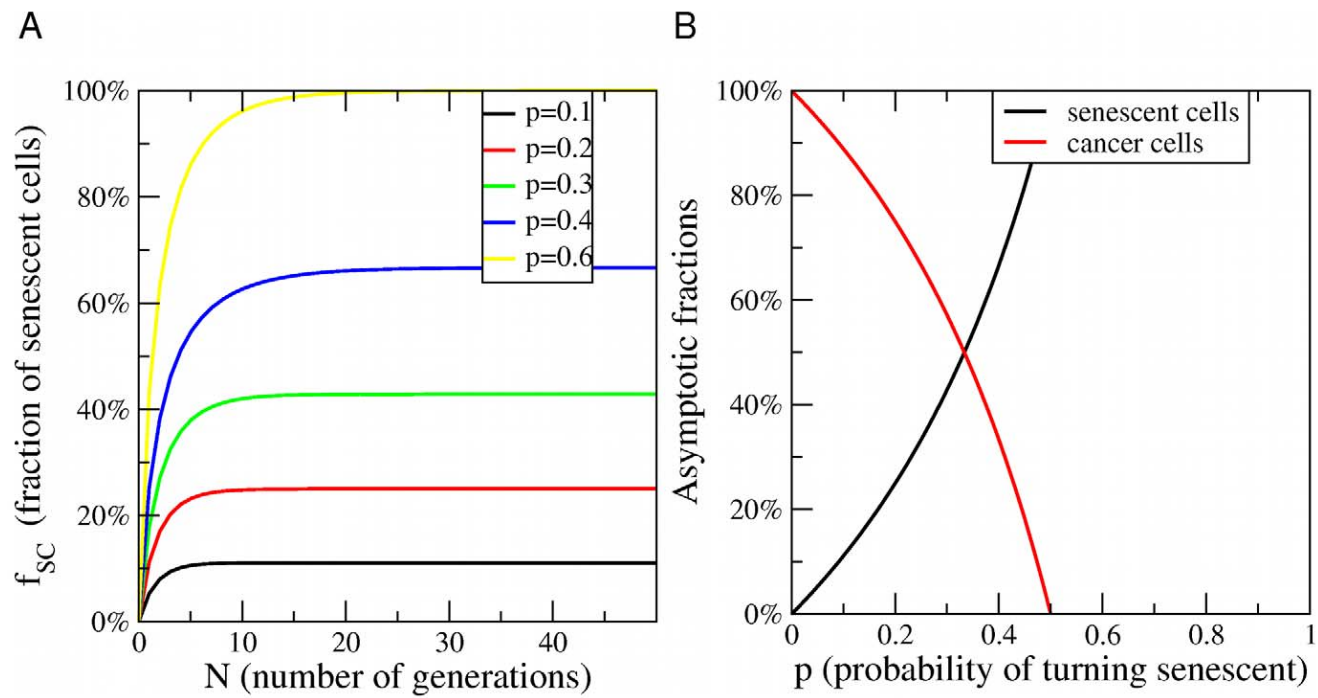

**Figure 6. Senescence in the stochastic model of cancer growth.** A) Evolution of the fraction of senescent cells as a function of the number of generations $N$ as a function of $p$ (compare with Fig. 2B for the CSC model) B) The asymptotic fractions of senescent cells and cancer cells as a function of the probability $p$ for a cell to become senescent. For $p>0.5$, all cells asymptotically become senescent and the tumor stops growing. (compare with Fig. 1B for the CSC model).
doi:10.1371/journal.pcbi.1002316.g006

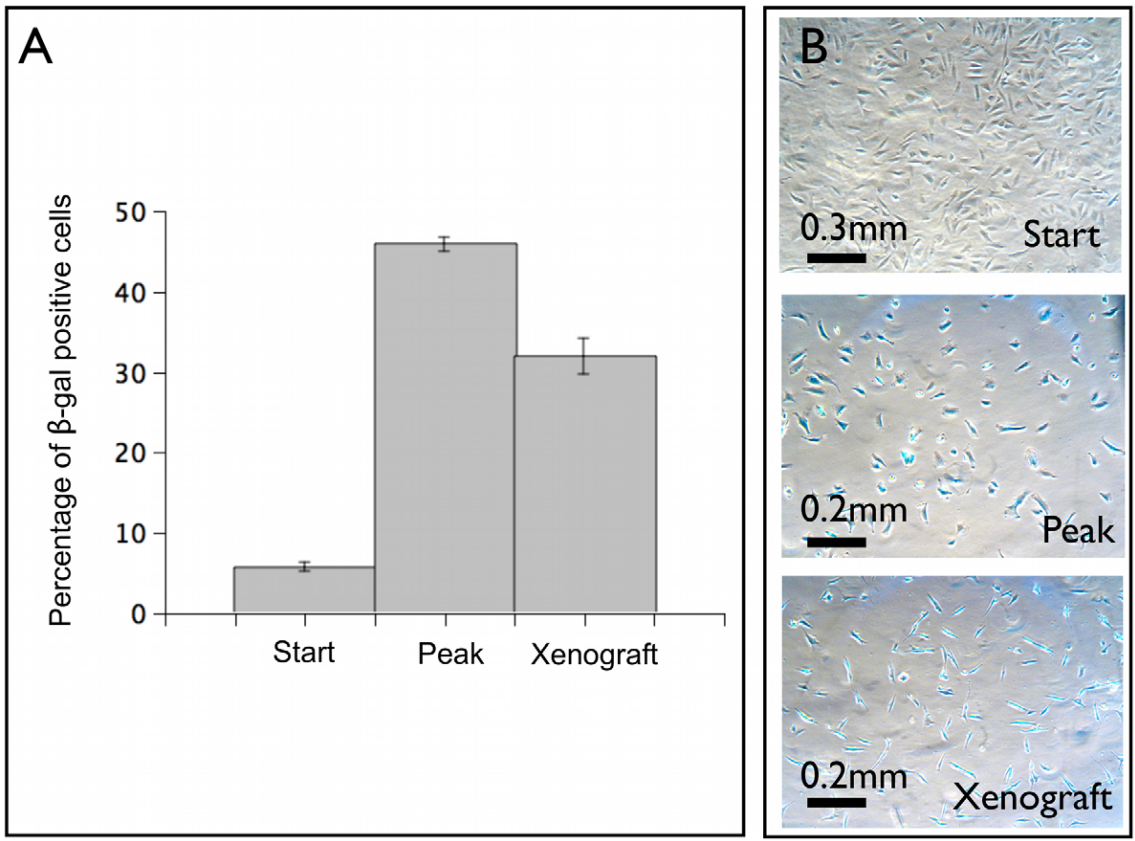

**Figure 7. Cellular senescence in human melanoma WM115 cells.** In panel A the presence of senescent cells in WM115 cells is quantified by the fraction of $\beta-$gal positive cells which we report at the beginning of the growth, at its peak and in tumor xenograft. The mean is taken over five independent determinations and the standard error is reported. Panel B shows $\beta-$gal staining (shown in blue) at the beginning (start), at peak and in tumor xenograft after 60 days.
doi:10.1371/journal.pcbi.1002316.g007

### Simulations of cancer treatments

It is illuminating to use the CSC model to simulate the effect of a treatment on the progression of a tumor. We consider two possible strategies: (i) try to stop tumor growth by stimulating cell senescence, and (ii) eradicate the tumor by inducing cell death. Case (i) can be described by decreasing the parameter $M$, representing the number of generation needed for a CC to turn senescent. The responses to treatment is summarized in Fig. 9A showing that the tumor size stops growing for a short while, but eventually the growth resumes at the previous rate. This is due to the action of CSC, whose fractional population increases dramatically in response to the treatment and sustains cancer growth. Our model quantifies the common-sense statement that if the cancer stem cells are the only parties that double forever, then a treatment that does not remove them will be fruitless in the long term. Case (ii) can be described by introducing a parameter $p$, as the probability that a CC cell dies (before senescence). If we assume that CSCs are drug resistant and therefore do not die, we predict that the tumor size would initially shrink, but the growth would starts again after some time (Fig. 9C), due to the persistent growth of the CSC subpopulation (Fig. 9D) Again, the model predicts that the only possible strategy to stop cancer growth is to target CSCs.

### Discussion

The present results have powerful implications for our picture of cellular senescence in tumors. The stochastic cancer model suggests that since cancer cells proliferate indefinitely, they must have evaded senescence by some yet unknown biological mechanism. Our results show instead that cancer cells in general become senescent, but this fact is ultimately irrelevant for tumor progression since long-term proliferation eventually resumes. This finding can be explained by the hierarchical cancer model where a subpopulation of CSCs drives tumor growth and the other cancer cells turn senescent after a fixed number of duplications. The peak in cell senescence should correspond precisely to the time at which the initial population of cancer cells turn senescent slowing tumor growth. At this point, CSCs are able to restart the growth process by symmetric duplication leading eventually to the decrease of the fraction of senescent cells as observed in the experiments. It would be hard to reconcile our experimental data with the conventional cancer model hypothizing that a random mutation would lead to massive senescence roughly at the same time for all the different cancer cell populations considered. The fact that CSCs do not become senescent is probably a signature that they originate from normal stem cells and are therefore already immortal. This observation may have implications for therapeutic strategies trying to stop tumor growth by inducing senescence: such an approach is bound to fail unless senescence is induced in CSCs.

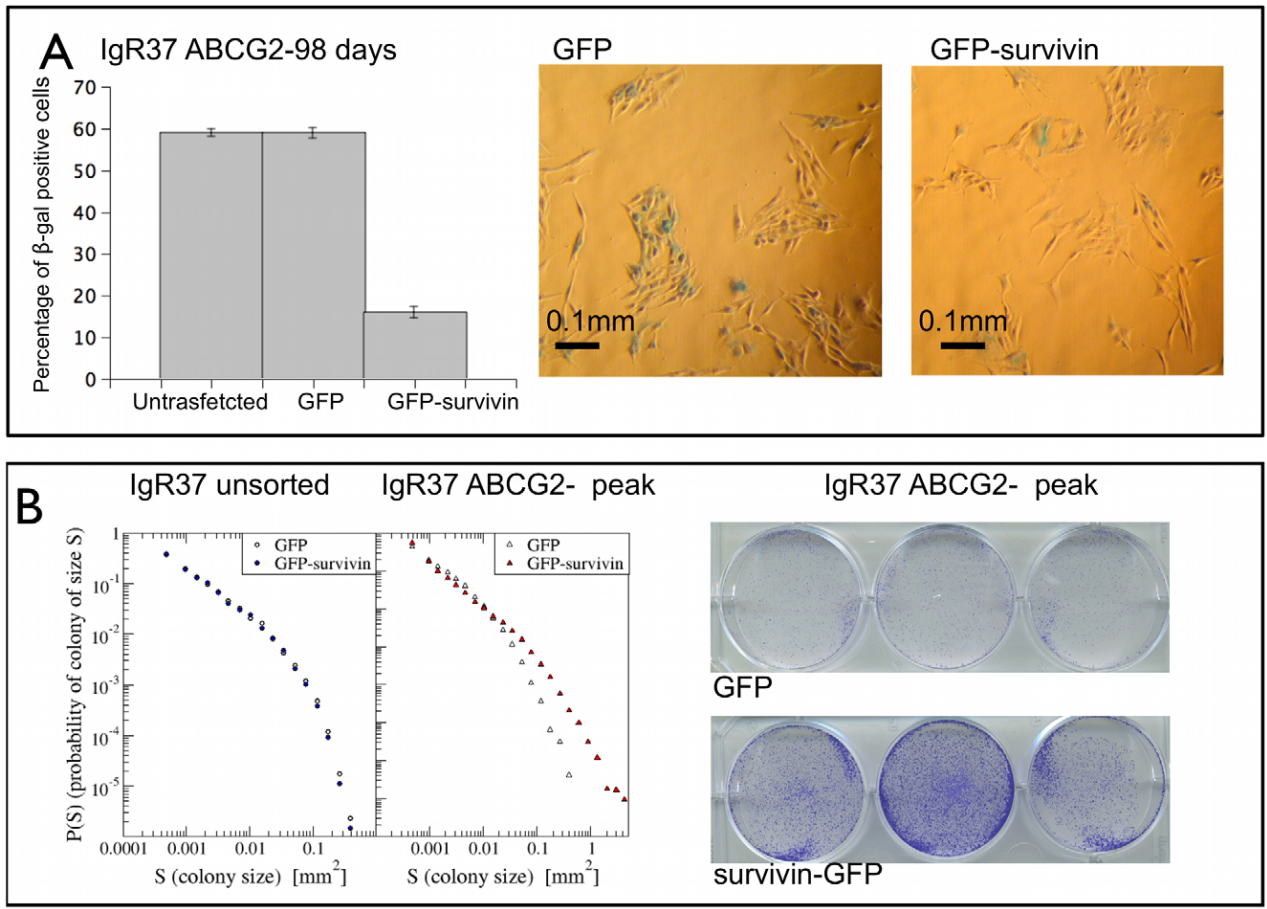

**Figure 8. Effect of survivin on senescence.** IGR37 ABCG2− cells were grown until the peak of $\beta$−gal (98 days see Table S1) and then transfected with a plasmid containing cytoplasmic-survivin-GFP or GFP alone. The next day the cells were plated for $\beta$−gal activity (50000cells/35mm² well) and crystal violet (200 cells per well) assays. In panel A, we show the percentage of $\beta$−gal positive cells for untransfected, GFP and GFP-survivin cells. The histogram shows the mean of five independent determination ± S.E. Represenative images of $\beta$−gal staining (shown in green) are reported for GFP and GFP-survivin cells. In panel B, we show the distribution of colony sizes from crystal violet assay for GFP and GFP-survivin cells, showing a marked increase in proliferation. Such an increase is not detected for unsorted IGR37 cells, transfected at the beginning of the experiment, for which the percentage of $\beta$−gal positive cells is already low.
doi:10.1371/journal.pcbi.1002316.g008

The data on the ABCG2 sorted population can be explained by the hierarchical cancer model if we assume that ABCG2+ and ABCG2− data are distinguished by a different initial fractions of CSCs. A larger fraction of CSCs in ABCG2+ cells should reduce the occurrence of senescence and increases proliferation with respect to ABCG2− cells. Furthermore, the observation that both cell populations eventually resume their growth at the same rate suggests that some CSCs are also present in ABCG2− cells, whose growth would otherwise stop. This implies that while ABCG2 is able to select a CSC rich population, by itself it is not a clean sorting criterion. This is in agreement with the results of Ref. [16] showing that ABCG2− cells yield a tumor xenograft in immunodeficient mice that is smaller than the one produced by ABCG2+ cells. The fact that ABCG2 is not an absolute CSC marker implies that its expression is reversible after sorting. We have checked this by sorting ABCG2+ and ABCG2− cells after 139 days finding that both express ABCG2 (with percentage 0.92% and 0.45%, respectively). The same problems are likely to affect putative CSC markers shown in the literature to be reversibly expressed and hitherto considered incompatible with the CSC model [18]. Similar observations in breast cancer cells have been recently interpreted as a signature of stochastic phenotypic switching, assuming that CC have a small probability to revert to the CSC state [19]. The same data could possibly be interpreted in our framework by assuming that the marker was not perfect. It would be interesting to develop quantitative tools to distinguish between phenotypic switching and imperfect markers.

The presence of CSC in melanoma is debated because conflicting data are reported in the literature [14,17,18] showing that slightly different assay conditions lead to different CSC fractions, that can sometimes be relatively large [17]. There is in fact no reason to believe that the CSC population must be small. This idea comes from the analogy with tissue stem cells that replicate homeostatically, keeping their population constant either by asymmetric division or stochastically [6,7], leading to a vanishing concentration of stem cells in the total cell population. CSCs do not replicate homeostatically and therefore their population grows exponentially. Changes in assay conditions can change the duplication rate of CSC, leading for extreme conditions (i.e. for example the use of matrigel, mice permissive conditions etc.) to a relatively large concentration of CSCs, perhaps resolving previous controversies [14,17,18].

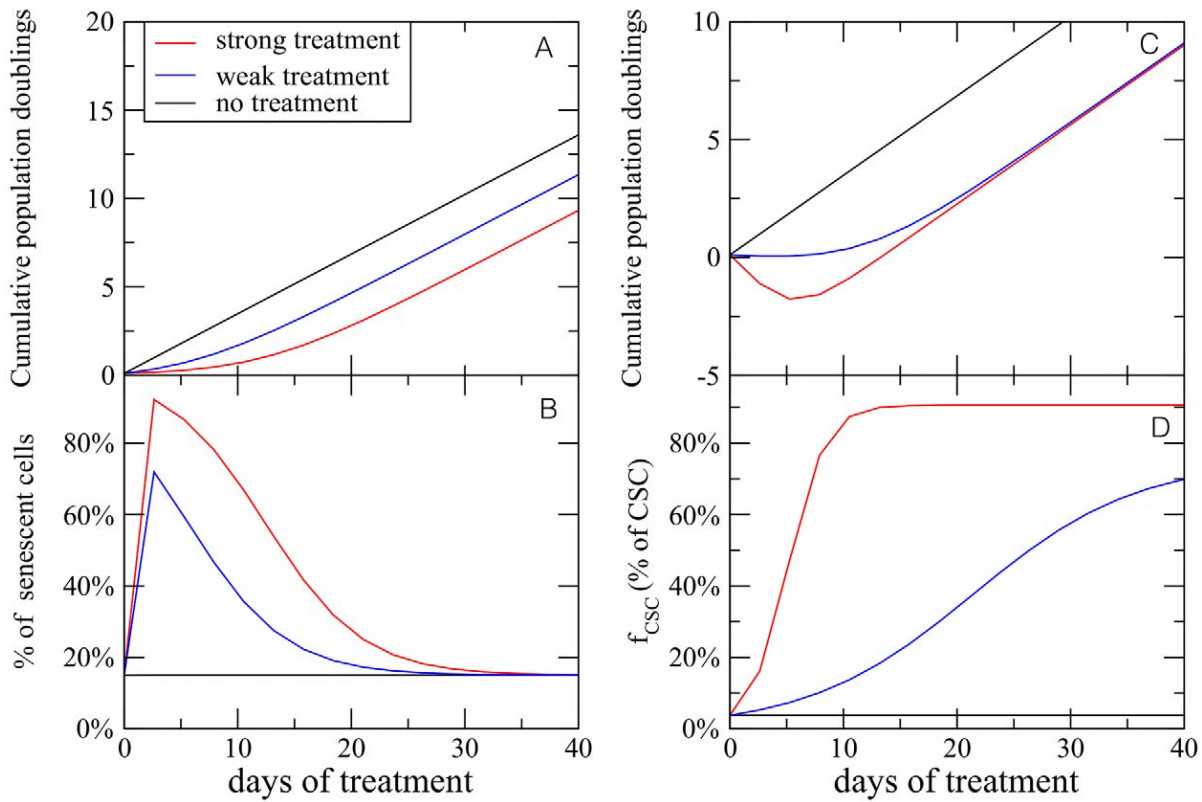

**Figure 9. Simulated response to drugs in the CSC model.** In panel A, we show the response of the model to a senescence inducing drug. The strength of the treatment is controlled by the parameter $M$, that in normal conditions we set to be $M = 20$. To simulate the effect of the drug, $M$ is reduced to $M = 10$ for a weak treatment and $M = 2$ for a strong treatment. The growth of the tumor is slowed down for a short period and then restart as before. Panel B shows that the fraction of drug induced senescent cells rapidly decreases to the level of the control. This is due to the CSC activity. In panel C, we show a simulation the effect of a cell death inducing drug. Here weak treatment corresponds to a probability $p = 0.2$ for a CC to die, while strong is simulated by setting $p = 0.7$. The drug leads to a rapid decrease of the tumor size, measured in terms of cumulative population doublings. After a few generations, however, the population starts to grow again. This is due to the CSC whose fraction increases as a response to the drug (panel D). These simulations are performed using parameters characteristic of IGR39 cells.
doi:10.1371/journal.pcbi.1002316.g009

## Materials and Methods

### Cell lines

Human IGR37 cells were obtained from Deutsche Sammlung von Mikroorganismen und Zellkulturen GmbH and cultured as previously described [16]. IGR39 was derived from a primary amelanotic cutaneous tumor and IGR37 was derived from an inguinal lymph node metastasis in the same patient. Human primary melanoma WM115 (ATCC CRL 1675) was cultured as in previous work [9].

### Flow cytometry

Cells were analyzed for fluoroscein isothiocyanate (FITC) mouse anti-human ABCG2 (R&D Systems, Minneapolis, MN) expression. All samples were analyzed using one- or two-color flow cytometry with non-specific mouse IgG used (Invitrogen, Carlsbad, CA) as isotype controls. For each flow cytometry evaluation, a minimum of $5 \times 10^5$ cells were stained and at least 50000 events were collected and analyzed ($10^6$ cells were stained for sorting). Flow cytometry sorting and analysis was performed using a FACSAria flow cytometer (Becton, Dickinson and Company, BD,

Mountain View, CA). Data were analyzed using FlowJo software (Tree Star, Inc., San Carlos, CA).

### Senescence associated $\beta$−gal assay

Expression of pH-dependent senescence associated $\beta$−gal activity was analysed simultaneously in different passage of cells using the Senescence galactosidase Staining Kit (Sigma) according to the manufacturers protocol.

### Tumor formation analyses and isolation of single cells from melanoma biopsy

$5 \times 10^5$ cells were injected subcutaneously into five-week-old NOD-SCID mice (Charles River Laboratories, Boston, MA). Animals were maintained on standard laboratory food ad libitum fed and free access to water. Tumor mass was excised after two months and digested with 0.2% collagenase type I (Gibco), 0.2% bovine serum albumin (BSA; SIGMA, St. Louis, MO) for 30 minutes at 37°C on an orbital shaker. Cells are maintained in culture for no more than two passages.

### Growth under hypoxic condition

WM115 cells were grown in standard medium under normal or hypoxic conditions (99.8% $CO_2$/0.2% $O_2$) for up to 48 hours.

Total RNA was extracted from WM115 cells or prostate cancer cell lines (PC-3 used as control) [32] under normal and hypoxic conditions. Reverse transcription (RT)-PCR reactions were done as previously described [16]. The oligonucleotide primer sequences and gene-specific PCR amplification programs used are following: sense 5′-CCTATGTAGTTGTGGAAGTTTATGC; antisense 5′-ACTAGGCAATTTTGCTAAGAATG for 40 cycles at 95°C for 30 s, 60°C for 15 s, and 72°C for 30 s. The integrity of each RNA and oligodeoxythymidylic acid-synthesized cDNA sample was confirmed by the amplification of the $\beta$-actin housekeeping gene. Ten microliter of each RT-PCR sample was run on a 2% agarose gel and visualized by ethidium bromide staining.

## Survivin overexpression

200000 cells were plated on 6 multiwell and the day after transfected with lipofectamin with cytoplasmic survivin-GFP plasmid or GFP (both plasmids were a gift of Dr Wheatley, [31]). 24 hours later the GFP-positive cells were counted by flow cytometry. More than 95% of cells were GFP-positive. These cells were immediately used for crystal violet or $\beta$-gal actosidase assays.

## Crystal violet staining cells and cluster analysis of the colonies

Cells were plated on 6 multiwell and stained after 8 days. Briefly, they were fixed with 3.7% paraformaldeide (PFA) for 5 minutes and then stained for 30 min with 0.05% crystal violet solution. After two washings with tap water, the plates were drained by inversion for a couple of minutes and then photographed. To quantify cell proliferation, we threshold the images and apply the Hoshen-Kopelman algorithm to identify individual clusters (see Fig. 4 for a visual explanation of the method). We then compute the number of colonies, the area fraction covered by colonies (see Table S2) and the colony size distribution (see Figs. 8B and S5). The last method is the most accurate and complete, avoiding possible experimental artifacts due to local cell detachments.

## Growth curves

Long-term growth curves are obtained by splitting the cells at each passage and rescaling the result of the counting by an appropriate factor. Cells are counted using a burken chamber in the presence of trypan blue. It is convenient to express the growth curves in terms of cumulative population doublings (CPD), expressing the number of times the populations has doubled. CPD is directly related to the number of cells $n_{tot}(t)$ at time $t$ by the expression:

$$CPD(t) = \log(n_{tot}(t)/n_{tot}(0))/\log 2; \quad n_{tot}(t) = n_{tot}(0)2^{CPD(t)} \quad (14)$$

where $n_{tot}(0)$ is initial number of cells.

## Fitting experiments with the CSC model

In order to fit the data with model, the total number of cells can be expressed as the sums of two terms $n_{tot}(t) = S^0 f(t) + c g(t)$ derived from Eq. 2 and Eq. 3, respectively. In particular, we have

$$f(t) = \left(\frac{2}{1+\epsilon}\right)^M (1+\epsilon)^{R_d t} \left[1 + \frac{1-\epsilon}{q+\epsilon}\left(1 - \left(\frac{1-q}{1+\epsilon}\right)^{R_d t - M}\right)\right], (15)$$

and

$$g(t) = \begin{cases} (R_d t - M)2^{R_d t} + (1-q)^{R_d t} + \dfrac{2}{1+q}(2^{R_d t} - (1-q)^{R_d t}) & \text{for } t \le M/R_d \\[2ex] (1-q)^{R_d t}\left[1 + \left(\dfrac{2}{1+q}\right)^M - \left(\dfrac{2}{1+q}\right)\right] & \text{for } t > M/R_d \end{cases}$$

$$(16)$$

Furthermore, we can also compare the model with the $\beta$-gal concentration, assuming that it is equal to the fraction of senescent cells. The experimental comparison involves the fit of four curves in terms of the parameters $M$, $R_d$, $\epsilon$, $q$ and $c$. Since ABCG2+ and ABCG2− cells only differ by their relative fraction of CSC, we use the same values of $M$, $R_d$, $\epsilon$, $q$ for the two subpopulations and vary only the parameter $c$. We thus perform a multiple curve fitting in which the two growth curves and the two $\beta$-gal curves are fitted simultaneously in terms of six parameters ($M$, $R_d$, $\epsilon$, $q$, $c^{(+)}$ and $c^{(-)}$), where $c^{(\pm)}$ is the value of $c$ for ABCG2± cells. To this end, we use the pyFitting software (https://github.com/gdurin/pyFitting) developed by G. Durin. The result is

$$R_d = 0.38 \pm 0.04 \quad (17)$$

$$M = 37.7 \pm 0.1 \quad (18)$$

$$\epsilon = 0.71 \pm 0.01 \quad (19)$$

$$q = 0.38 \pm 0.04 \quad (20)$$

$$c^{(+)}/S^0 = 0.13 \pm 0.01 \quad (21)$$

$$c^{(-)}/S^0 = 4.1 \pm 0.3 \quad (22)$$

with a reduced $\chi^2$ of 14.3. In Table S3, we report the covariance matrix and the t-statistic.

We can express the parameters $c^{(+)}$ and $c^{(-)}$ in terms of the initial fractions of CSC as $f_{CSC}^0 = S^0/(S^0 + c^{(\pm)}(M+1))$ for positive and negative cells, where the arbitrary scale $S^0$ is set equal to one in the fits. We obtain

$$f_{CSC}^0 = 0.63 \pm 0.05\% \text{ for ABCG2}{-} \quad (23)$$

$$f_{CSC}^0 = 16 \pm 1.\% \text{ for ABCG2}{+} \quad (24)$$

Hence the ABCG2+ cells have more CSCs than ABCG2− cells, as expected. We have also checked how much the choice of the initial condition affect the results. For simplicity, we have imposed a uniform distribution of $C_k^0$ (for $k > 0$) to obtain a closed form solution (Eqs. 2–3). To assess the robustness of our results, we have compared the growth curves obtained under this assumption with those obtained using the same fit parameters and $C_k^0$ imposed as the steady-state solution of the model with the same parameters but smaller $M$. The resulting curves are very close to each other indicating that at least for this case there is a weak dependence on the initial conditions.

## Supporting Information

**Figure S1  Asymptotic fractions of CSCs and senescent cells in the model.** The asymptotic fraction of CSCs (A) and senescent cells (B) as a function of the proliferation parameter $\epsilon$ and for different values of the number $M$ of duplications needed by cancer cells to become senescent and of the probability of cell death $q$. Unlike the fraction of CSC, the fraction of senescent cells does not depend on $M$.
(TIFF)

**Figure S2  Growth curves of mesenchymal stem cells (HIP).** The growth of populations of MSC isolated from the bone marrow from the femoral hip in terms of cumulative population doublings are fitted by the model. Experimental data are obtained from Ref. [29]. Cell populations refer to donors with different ages. The best fit is obtained varying $M$ and $R_d$ in Eq. 7.
(TIFF)

**Figure S3  Hypoxia inducible factor.** RT-PCR of $HIF-1\alpha$ (145bp) of WM115 cells grown under normal or hypoxic condition (from 18 to 48 hrs) and PC-3 cells as positive control.
(EPS)

**Figure S4  Cluster analysis of crystal violet assay.** A) To analyze the crystal violet assay, we first photograph the multiwell and isolate a single well. B) Using the image analysis software Gimp we select by color the spots, eliminate the background and threshold the remaining spots in order to obtain a two color image. C) We apply the Hoshen-Kopelman cluster algorithm to identify individual colonies, recolored here with random colors for visualization purposes.
(TIFF)

**Figure S5  Effect of survivin on senescence away from the peak of $\beta-gal$.** we show the distribution of colony sizes

obtained from crystal violet assay for untransfected, GFP and GFP-survivin cells. In this experiment 500 cells were plated after 88 days of cultivation (see Table S2). We see a small effect due to GFP and survivin.
(TIFF)

**Table S1  Evolution of senescence marker.** The evolution of the percentage of $\beta-gal$ positive cells for ABCG2 sorted IGR39 and IGR37 cell populations.
(PDF)

**Table S2  Crystal violet assays.** A summary of the results obtained with the crystal violet assay for GFP and GFP-survivin transfected cells at different stages of the growth. We report the total number of colonies and the fraction of area covered by colonies.
(PDF)

**Table S3  Fit covariance matrix.** The covariance matrix and the t-statistics of the joint fit of the growth curves and $\beta-gal$ concentration with the CSC model.
(PDF)

## Acknowledgments

We thank Emilio Ciusani (Istituto Neurologico Besta, Milano, Italy) for cell sorting, Susan Wheatley (University of Sussex, Brighton, UK) for providing Survivin-GFP plasmid and Gianfranco Durin (INRIM, Torino, Italy) for help with curve fitting. Finally, we are grateful to James L. Sherley (BBRI, Watertown, MA, USA) for helpful discussions and suggestions.

## Author Contributions

Conceived and designed the experiments: CAMLP. Performed the experiments: CAMLP. Analyzed the data: SZ. Wrote the paper: CAMLP SZ. Designed and solved theory: SZ JPS.